\documentclass[letterpaper, 10 pt, conference]{IEEEtran}
\IEEEoverridecommandlockouts
\usepackage{cite}
\usepackage{amsmath,amsfonts, amsthm}
\usepackage{algorithmic}
\usepackage{graphicx}
\usepackage{textcomp}
\usepackage{xcolor}
\usepackage{hyperref}
\usepackage{subcaption}
\usepackage{svg}
\theoremstyle{definition}
\newtheorem{definition}{Definition}

\def\BibTeX{{\rm B\kern-.05em{\sc i\kern-.025em b}\kern-.08em
    T\kern-.1667em\lower.7ex\hbox{E}\kern-.125emX}}

\begin{document}

\title{\vspace{0.6cm}Prototyping Autonomous Robotic Networks on Different Layers of RAMI 4.0 with Digital Twins}

\author{
\IEEEauthorblockN{Alexander Barbie\IEEEauthorrefmark{1}\IEEEauthorrefmark{2}\IEEEauthorrefmark{3}, Wilhelm Hasselbring\IEEEauthorrefmark{3}, Niklas Pech\IEEEauthorrefmark{2}, Stefan Sommer\IEEEauthorrefmark{2}, Sascha Fl\"ogel\IEEEauthorrefmark{2}, Frank Wenzh\"ofer\IEEEauthorrefmark{1}}
\vspace{0.3cm}
\IEEEauthorblockA{\IEEEauthorrefmark{1}Alfred Wegener Institute Helmholtz Centre for Polar and Marine Research (Germany)}
\IEEEauthorblockA{\IEEEauthorrefmark{2}GEOMAR Helmholtz Centre for Ocean Research Kiel (Germany)}
\IEEEauthorblockA{\IEEEauthorrefmark{3}Software Engineering Group, Christian-Albrechts-University, Kiel (Germany)}
}

\maketitle

\begin{abstract}
In this decade, the amount of (industrial) Internet of Things devices will increase tremendously. Today, there exist no common standards for interconnection, observation, or the monitoring of these devices. In context of the German ``Industrie 4.0'' strategy the Reference Architectural Model Industry 4.0 (RAMI 4.0) was introduced to connect different aspects of this rapid development. The idea is to let different stakeholders of these products speak and understand the same terminology. In this paper, we present an approach using Digital Twins to prototype different layers along the axis of the RAMI 4.0, by the example of an autonomous ocean observation system developed in the project ARCHES.
\end{abstract}


\section{INTRODUCTION}
During the past decade, a rapid development of new technologies paved the way for the (industrial) Internet of Things (IoT). IoT devices are embedded systems, that are connected to the Internet or operate in a local area network. This cross section of physical objects and the processing with computer systems is called a Cyber-Physical System (CPS) and is a common industrial engineering practice \cite{cps}. Today, CPS are not only used for smart home devices, but also for manufacturing plants, cars, and in many other domains. In Germany's economy the strategy ``Industry 4.0'' is an umbrella term for (industrial) IoT and CPS, and describes the fourth industrial revolution.

\begin{figure}[tb]
  \centering
  \includegraphics[width=0.49\textwidth]{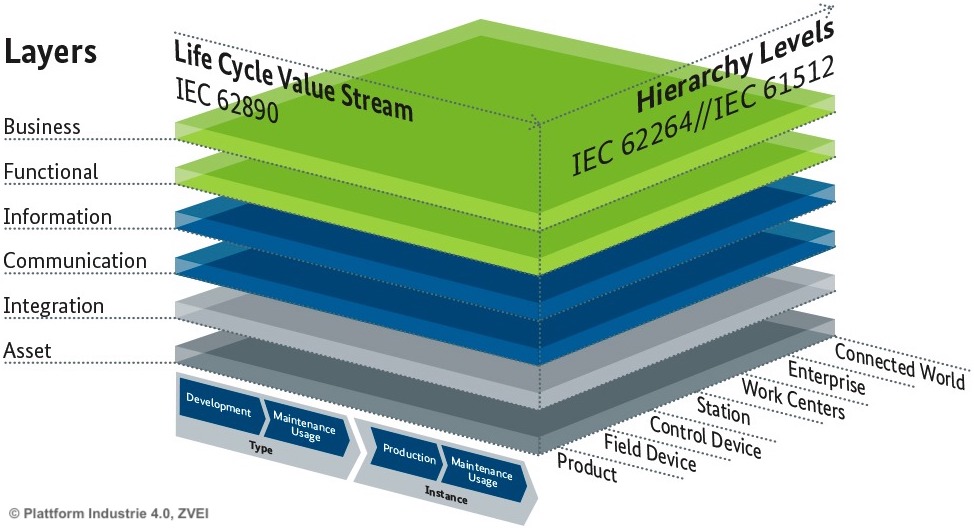}
  \caption{Reference Architectural Model Industry 4.0 \cite{plattform40}}
  \label{fig:rami4.0}
\end{figure}

One major challenge for Industry 4.0 is the very large amount of different devices without common standards for interconnection, observation, and monitoring.
In $2015$ the German Commission for Electrical, Electronic and Information Technologies of DIN and VDE (DKE) published the Reference Architectural Model Industry 4.0 (RAMI 4.0), which includes the technical framework (ISA-95) and standards (IEC62264/61512/62890) \cite{plattform40, ramizvei}. The RAMI layered architecture is shown in \autoref{fig:rami4.0}.
RAMI $4.0$ advocates a service-orientated architecture and is a three-dimensional model along three axes (architecture, product life cycle, and hierarchy) \cite{cps}. The dependency between layers is downward, i.e., the change of an upper layer has no effect on the layers below.

In the context of Industry $4.0$ the term ``Digital Twin'' (DT) became a buzz-word and refers to a virtual copy of a physical product \cite{grieves2011virtually,cps}. Originally, the ``Twin'' concept was used by the NASA in the late $1960$s during the Apollo missions and was developed further to ``Digital Twin'' in $2010$ \cite{nasadt2012}. However, there still is no common definition for DTs and hence, they differ from simulation only to digital copy of a physical thing with state-synchronization in real-time, always depending on the domain where the DT concept is used.

In this paper, we follow the DT approach to develop Ocean Observation Systems (OOS) as autonomous robotic networks. By this example we emphasize the advantages of this approach in developing different layers along the different axis on the RAMI 4.0. 

In \autoref{sec:relatedwork} the previous and related work is presented and we show which contributions we offer to the DT concept. The DT concept used in this paper is defined in \autoref{sec:dt}. Our approach of using the prototyping technique to develop parts on different layers of the RAMI 4.0 is presented in \autoref{sec:sedt}. Some limitations of this approach are discussed in \autoref{sec:threats}. The conclusion and future work is presented in \autoref{sec:conclusion}

\section{PREVIOUS AND RELATED WORK}\label{sec:relatedwork}
The presented work takes place within the project ARCHES (Autonomous Robotic Networks to Help Modern Societies) funded within the Helmholtz framework ``Zukunftsthemen'' involving partners from the German Aerospace Center (DLR), Karlsruhe Institute of Technology (KIT), the GEOMAR Helmholtz Center for Ocean Research Kiel, and the Alfred-Wegener-Institute (AWI). 
One major goal of this project is to establish a sensing underwater network of heterogeneous, autonomous and interconnected robotic systems. The network consists of different stationary and mobile sensing platforms partly expands on measurement platforms developed within the HGF funded project ROBEX (Robotic Exploration of Extreme Environments) \cite{robex, wenzhofer2016tramper, robex2018}. However, there is no common interface to connect the robotic systems in an underwater acoustic network. Please note, that techniques depending on electromagnetic waves cannot be used, since they are absorbed by sea water.

In the past, embedded systems that operate in extreme environments were tested by building identical duplicates and simulating different scenarios with these duplicates. The NASA first started with this concept within the Apollo program, where various identical space vehicles were built. During these space missions, one vehicle, called the ``Twin'', remained on earth and mirrored the conditions of the vehicle in space \cite{importancedt}. In future NASA missions, the Twin is a virtual representation of the vehicle in space called a ``Digital Twin'' \cite{nasadt2012}. Another example of a complex simulation of the interaction between hard- and software is the Iron Bird by Airbus. Airbus created a skeleton of their aircrafts A350 and A380 in a test rig, containing all components except the jet engines,  seats, and outer casing. Using such kind of hardware twin, they tested the flawless operation of the electric, hydraulic, flight control, and trained pilots in a corresponding flight-simulator \cite{briere1993airbus}. This test bench represents a good example for Hardware-in-the-Loop (HiL) tests \cite{hil1999} at a very large scale. However, in literature many different definitions of a DT can be found \cite{importancedt, nasadt2012, dtdefinition}. For example, another approach to implement DTs in an industrial context is presented in \cite{kafkadt}. The authors use a microservice-based Apache Kafka stream domain-specific language to implement DTs and emphasize the advantage of automatic testing their software with this concept. For a survey of prototyping software systems, refer to \cite{ACMCS2000}.

\section{DEFINING THE DIGITAL TWIN}\label{sec:dt}
For consistency, we defined the DT and its part in this paper, based on the DT definition provided by Roberto Saracco \cite{dtdefinition}, as follows:

\begin{definition}[Digital Model]
A digital model describes a real entity such as an object, a process, or a complex aggregation. The description is either a mathematical or a computer-aided design (CAD) model.
\end{definition}

Most definitions of digital models contain only the CAD model. We use a broader interpretation, since our use case is not only to monitor the physical parts of an OOS. Our focus is on the state of internal parts and the environmental data the system is operating in.

\begin{definition}[Digital Shadow]
A digital shadow is the sum of all the data that are gathered by an embedded system from sensing, processing, or actuating.
\end{definition}

Everyone who uses Internet services on a mobile phone or computer leaves a digital shadow behind that is traceable and closely linked to the user. Today, there is a whole economy that builds on using these digital shadows to sell products. This analogy is applied to physical objects that produce data with every action they perform. 

\begin{definition}[Digital Thread]\label{def:digitalthread}
The digital thread refers to the communication framework that allows a connected data flow and integrated view of the physical twin's digital shadow throughout its lifecycle.
\end{definition}

A digital shadow of physical objects can only be persisted, if the objects have an interface that stores the emerging data. The digital thread on mobile devices is the Internet. For digital twins there is no universal and generally valid solution right now, due to the different domains digital twins are used in.

\begin{definition}[Digital Twin]
A digital twin is a digital model of a real entity, the physical twin. It is both a digital shadow reflecting the status/operation of its physical twin, and a digital thread, recording the evolution of the physical twin over time.
\end{definition}

When testing new software on a digital twin that is connected to its physical twin, the number of developers that can test this software at the same time is limited to one. Therefore, every developer needs an instance of the digital twin independently of his/her access to the physical twin. This can be achieved by using a digital shadow from past missions that replays the behavior of a physical twin, although the instance is not connected to the real system.

\begin{definition}[Digital Twin Prototype]
A digital twin prototype (DTP) is the software/model prototype of a real entity, the physical twin. It uses existing recordings of sensing and actuation data over time as digital shadow, to simulate the physical twin. Often the DTP exists before there is a physical twin.
\end{definition}

Today, it is a common practice to create CAD models before building the physical prototype. In this paper, the DTP is a software prototype.

\section{PROTOTYPING IN RAMI 4.0}\label{sec:sedt}
In a simple single robotic sensing platform, which harbors several sensors and stores data locally on-board, sensors can easily be added, removed, and replaced without touching the operational software of this platform. This approach is stretched to its limits the moment a researcher or another instance needs to access or process the data of a sensor during run-time. To access the sensor and retrieve its data, an additional system with access to the sensor is needed. When adding or removing sensors, this system has to be adapted to the changes. With each additional requirement the complexity of these observation systems increases. The operating software of commercially available Remote-Operating-Vehicles (ROV) and Autonomous-Underwater-Vehicles (AUV) often represent a black box and has a monolithic architecture. It is not possible to add a new sensor without major changes of the code. Especially in the field of embedded systems, changing software is difficult and prone to failure. 

\subsection{Along the Layer Axis of RAMI 4.0}\label{subsec:dtp}
The following approach is presented by the example of a single OOS that serves as an asset. Nevertheless, this approach is used for all other OOSs in the project ARCHES, too.
\subsubsection{Asset}
The asset in this work is a stationary bio-geochemical OOS, the BIGO Lander \cite{bigo2003}. This stationary platform is an instrument carrier system and is used to study processes at the benthic boundary layer. Usually, it is deployed on the seafloor at depths of several hundred to $6,000$ meters for up to twelve months \cite{bigo2003}. These systems are released from a research vessel and reach the seafloor in a free-fall mode. The instruments that take bio-geochemical samples are connected to control units that operate with fixed routines to measure every several hours. All data are stored on-board. Only after the OOS is retrieved from the seafloor, researchers can access the data.

With further development in ocean science, new scientific questions and methods emerged and researchers want to access the gathered data as fast as possible, not after several months. Since the BIGO Lander is mostly used in other contexts than the project ARCHES, the software on its  control units remains untouched. However, its control units have RS232 interfaces to exchange commands and data. Hence, we added a RaspberryPI 3B+ that runs our software and is connected to the control units to operate them.

\subsubsection{Integration} To be able to change sensors carried by a OOS without major code changes, a modular embedded software system with loosely coupled nodes is required. Additionally, the OOS is expected to react to environmental changes or temporary events such as storms. These kinds of systems, are assumed to properly work based on an Event-Driven Architecture (EDA). An EDA consists of different microservices that exchange data via events. In this approach, an EDA is implemented using the Robot Operating System (ROS), which is no complete operation system, but a middleware that may be installed on Linux \cite{rospaper}. The advantage of ROS over other frameworks is the large open-source community that steadily improves the framework. It provides all required interfaces and services such as a publish-subscribe service, to develop a microservice architecture with loosely coupled nodes with low inter-dependencies. The distributed nodes use TCP connections to transfer ROS messages. These messages are defined like a schema in separate text files and are compiled to program code. The different fields in the schema use C primitive types. Each node either subscribes to a topic of events or publishes an event to a topic. We use Docker \cite{docker} as a containerization platform, to encapsulate the different microservices.

\begin{figure*}[tb]
  \centering
  \includegraphics[width=0.9\textwidth]{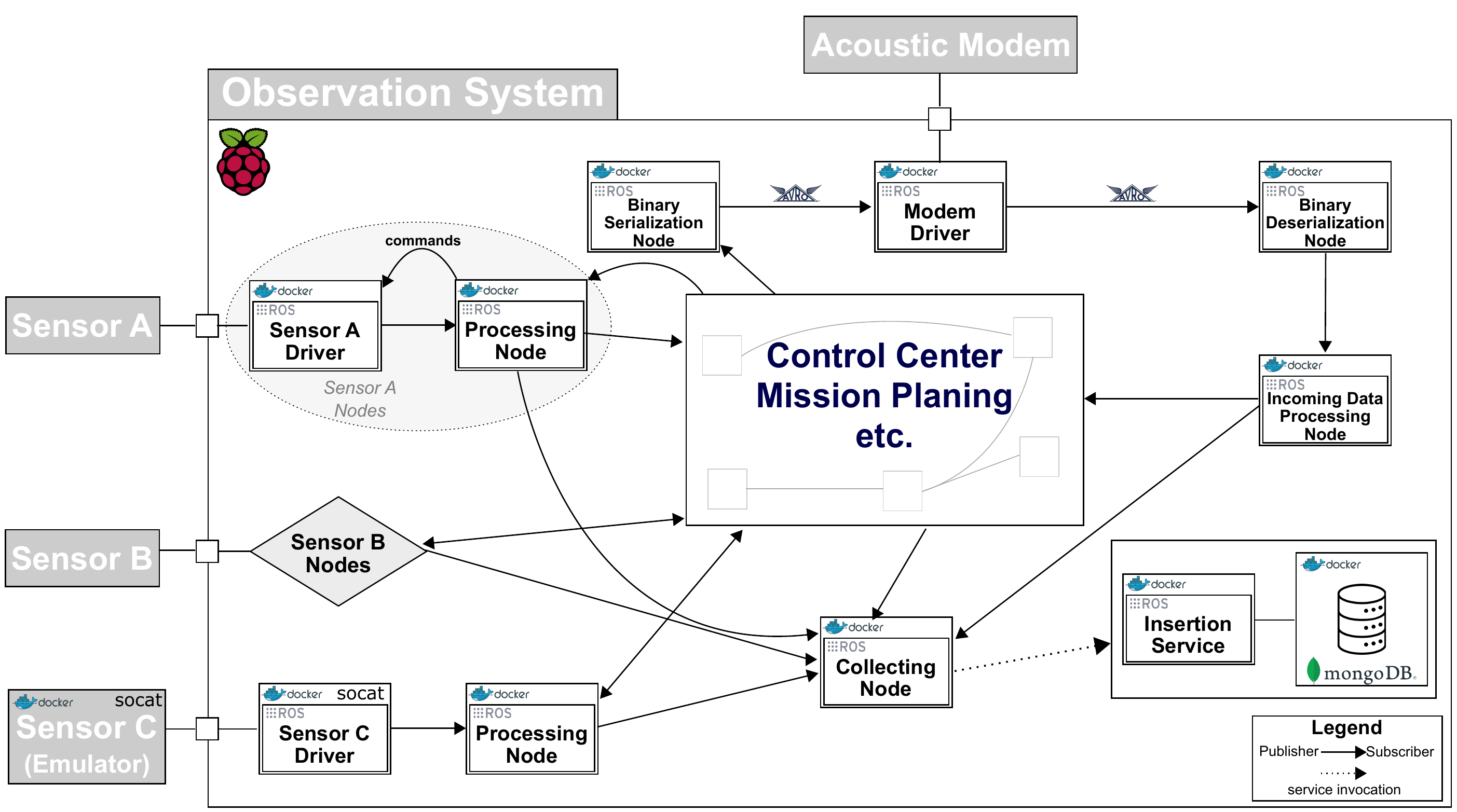}
  \caption{The microservice architecture of the DTP.}
  \label{fig:architecture}
\end{figure*}

\begin{figure}[tb]
  \centering
  \includegraphics[width=0.45\textwidth]{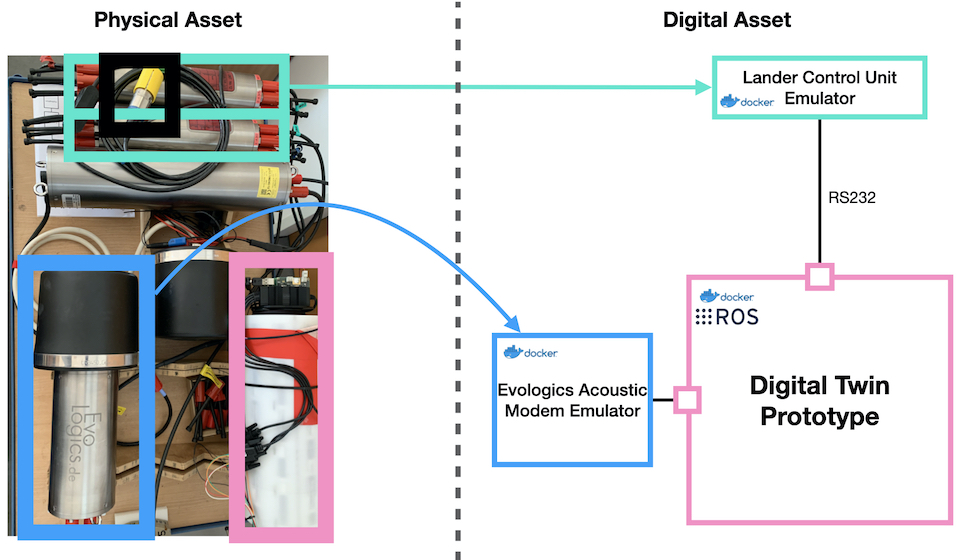}
  \caption{From test rig to a Digital Twin}
  \label{fig:bigotestrig}
\end{figure}

On the left side of \autoref{fig:bigotestrig} a test rig of a BIGO Lander is shown. This test rig contains two Evologics acoustic modems \cite{evologicspage} (blue box), two Lander Control Units (LCU) (turquoise boxes), one oxygen sensor (Aanderaa O2 Optode, black box) that is connected to a LCU, and a RaspberryPI 3B+ (RPi) with a battery pack (pink box). 

The first step towards a DTP of the BIGO Lander, as shown on the right side of \autoref{fig:bigotestrig}, is the development of emulators of the different components. At this point it is crucial that the formats of the in- and outputs of the emulators are identical to the formats used by the sensors. This way, the virtual representation of the important parts of the software system is created and enables developers to incrementally extend the system, which is the purpose of prototyping, without the need of a physical connection to the system. For the acoustic modems, an emulator was commercially available from Evologics (Berlin). For the LCU a Python program that waits for commands via the RS232 interface and responds with data strings was developed. This program is executed inside a Docker container. On the DTP, a ROS node was developed to connect to the LCU via the given RS232 interface and send commands or receives data. The ROS node is executed inside a Docker container, too. 

To simulate the RS232 interface between the emulator and the ROS node, Socat \cite{socat}, a tool for data transfer between two addresses, is used. An address can represent a network socket, TCP and UDP (over both IPv4 and IPv6), PTY, OpenSSL, and many more. For each Docker container, a customized bash script is used for its entry point. This script connects to a virtual TTY via Socat, if environmental variables are set on start-up of a container. Otherwise, the real TTY socket is passed to the container. This way, a DTP can either connect to the test rig and execute everything on the real hardware, or use the emulators to test the behavior. Furthermore, according to the idea of the RealPeer approach for simulation-based development \cite{realpeer}, work with a mix of real and emulated hardware can be conducted.
The resulting architecture is shown in \autoref{fig:architecture}.

\subsubsection{Communication} The advantage of the approach presented in this paper is not only the independence of the connection to a physical asset. It also enables developers to incrementally add new sensors and actuators. Above all, it enables developers to start additional instances. Such additional instances are very useful for prototyping the digital thread, which is an essential part of the DT concept. One instance uses the emulators to generate data and to send them to the other instance, which receives the data and processes it. Using this method, we develop the synchronization of the physical twin with the digital twin.

However, in an industrial context the connection would be some kind of a TCP/IP connection with a high bandwidth. The digital shadow then contains as many information as possible from the product. The challenge is to analyze the gathered data in real time and filter the most important information in a human readable way \cite{henning2019scalable}. This is not possible in underwater communication, where acoustic modems only have a limited bandwidth. In our context we are limited to $64$ bytes per second or per message to send. Thus, the Evologics emulator was used to create a protocol with a binary de-/serialization of ROS messages using Apache Avro$^{TM}$ \cite{avro}. This binary de-/serialization is important, since ROS messages are serialized to JSON strings that have a very large overhead of characters.

\begin{figure*}[tb]
\centering
\begin{subfigure}{0.34\textwidth}
\centering
  \includegraphics[width=\textwidth]{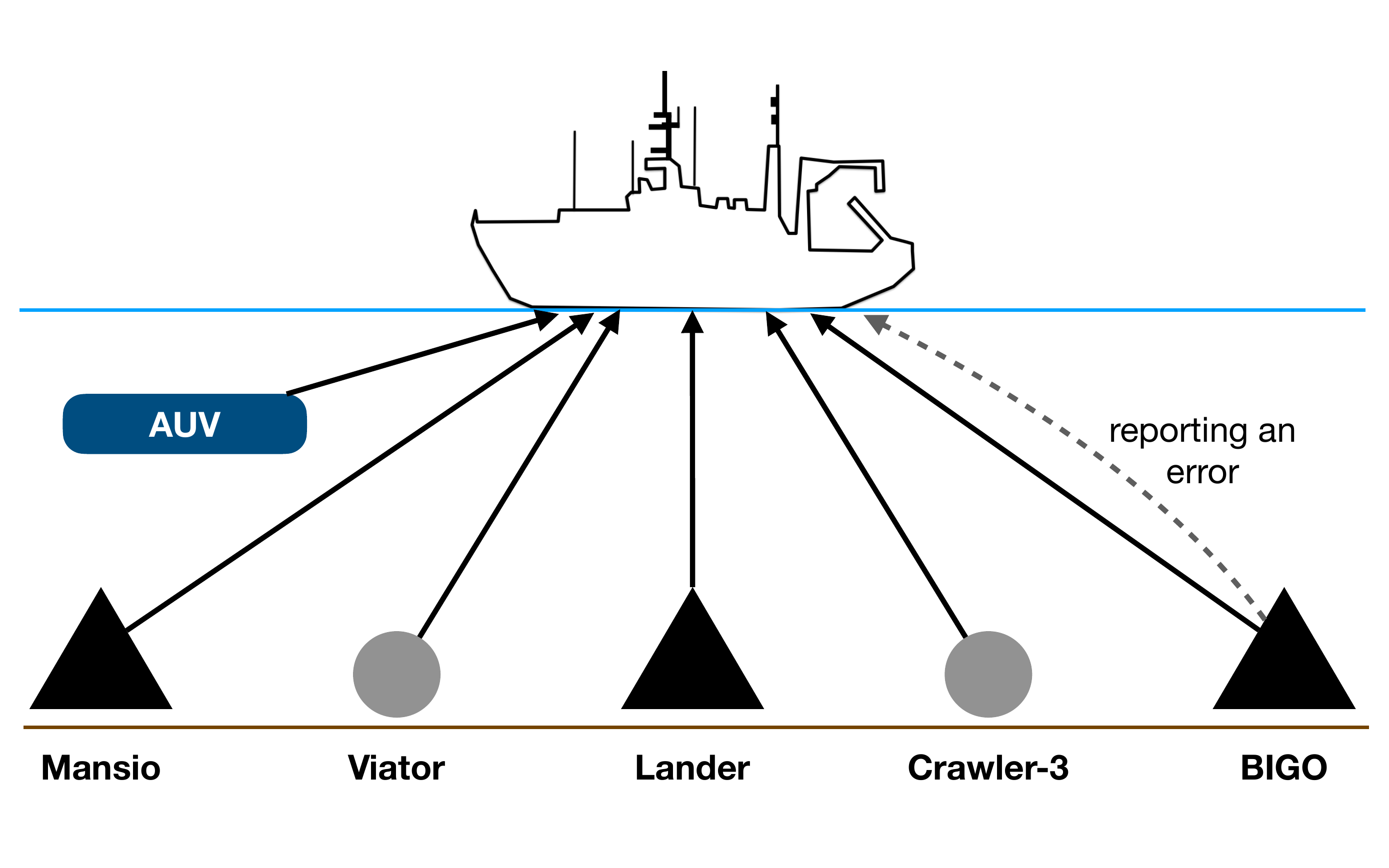}
  \caption{Send data to the research vessel}
  \label{fig:scenarioa}
\end{subfigure}
\hspace{2.5cm}
\begin{subfigure}{0.34\textwidth}
\centering
  \includegraphics[width=\textwidth]{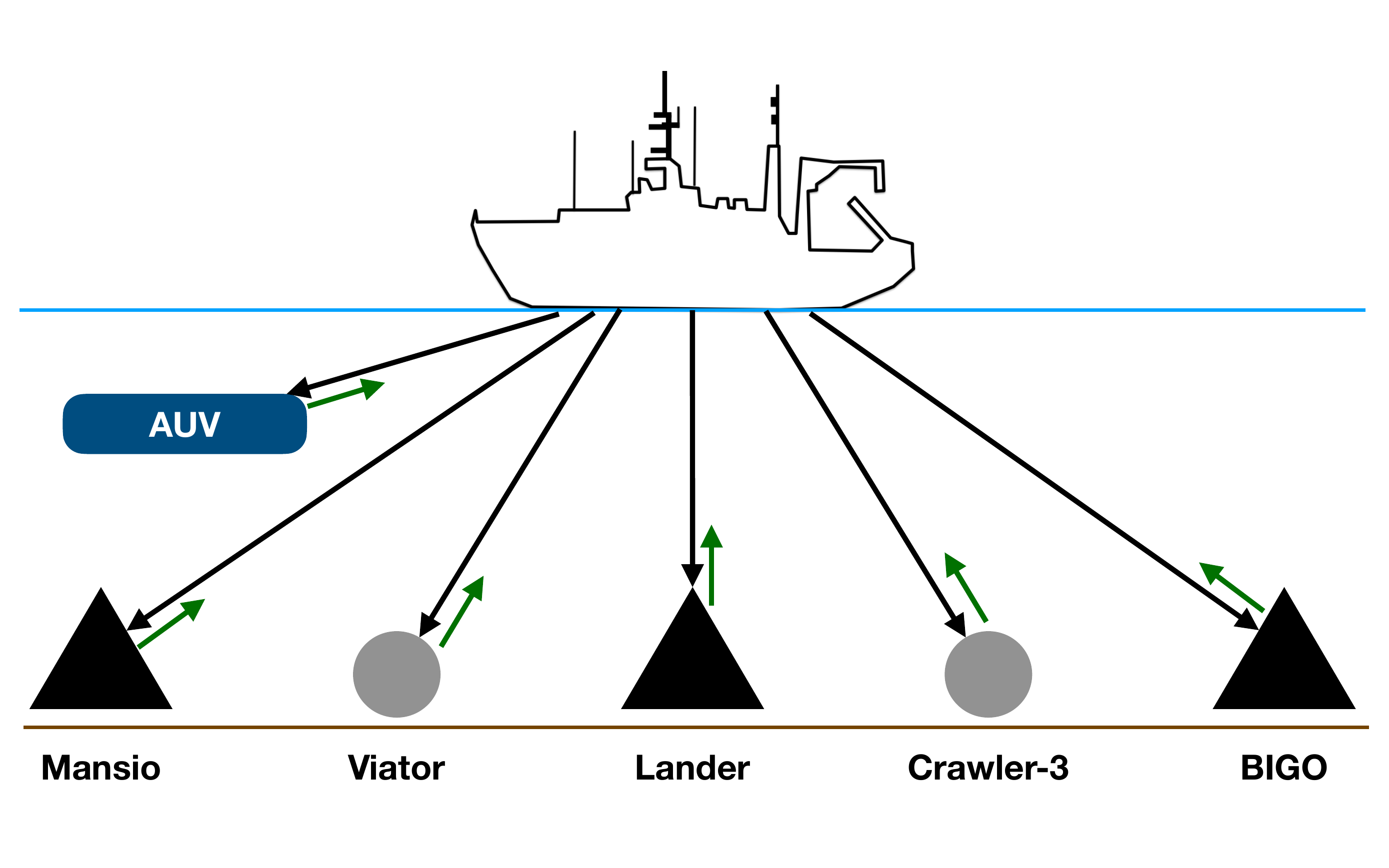}
  \caption{Send commands from the research vessel}
  \label{fig:scenariob}
\end{subfigure}
\\
\begin{subfigure}{0.34\textwidth}
\centering
  \includegraphics[width=\textwidth]{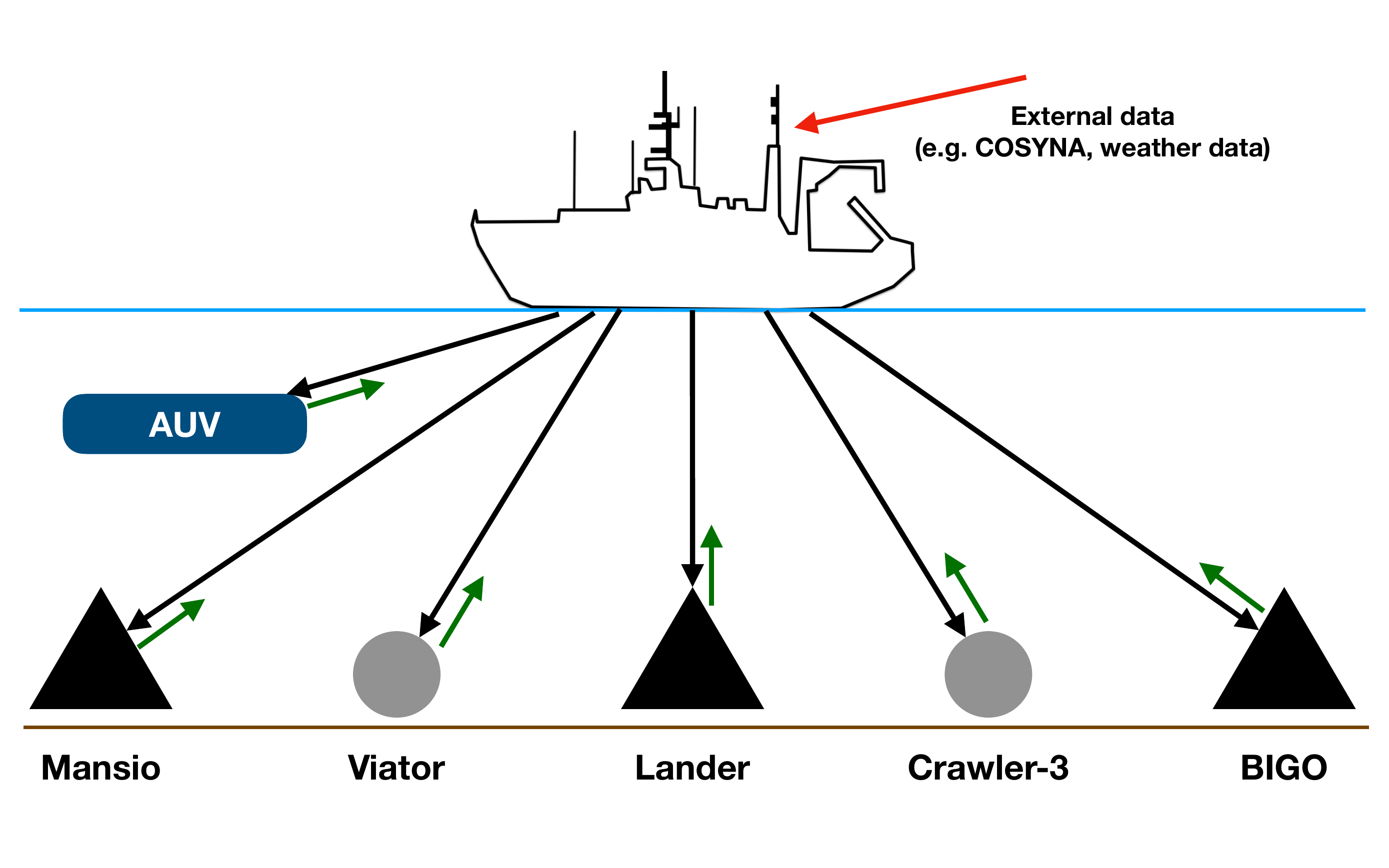}
  \caption{Use external data and send an event from the research vessel}
  \label{fig:scenarioc}
\end{subfigure}
\hspace{2.5cm}
\begin{subfigure}{0.34\textwidth}
\centering
  \includegraphics[width=\textwidth]{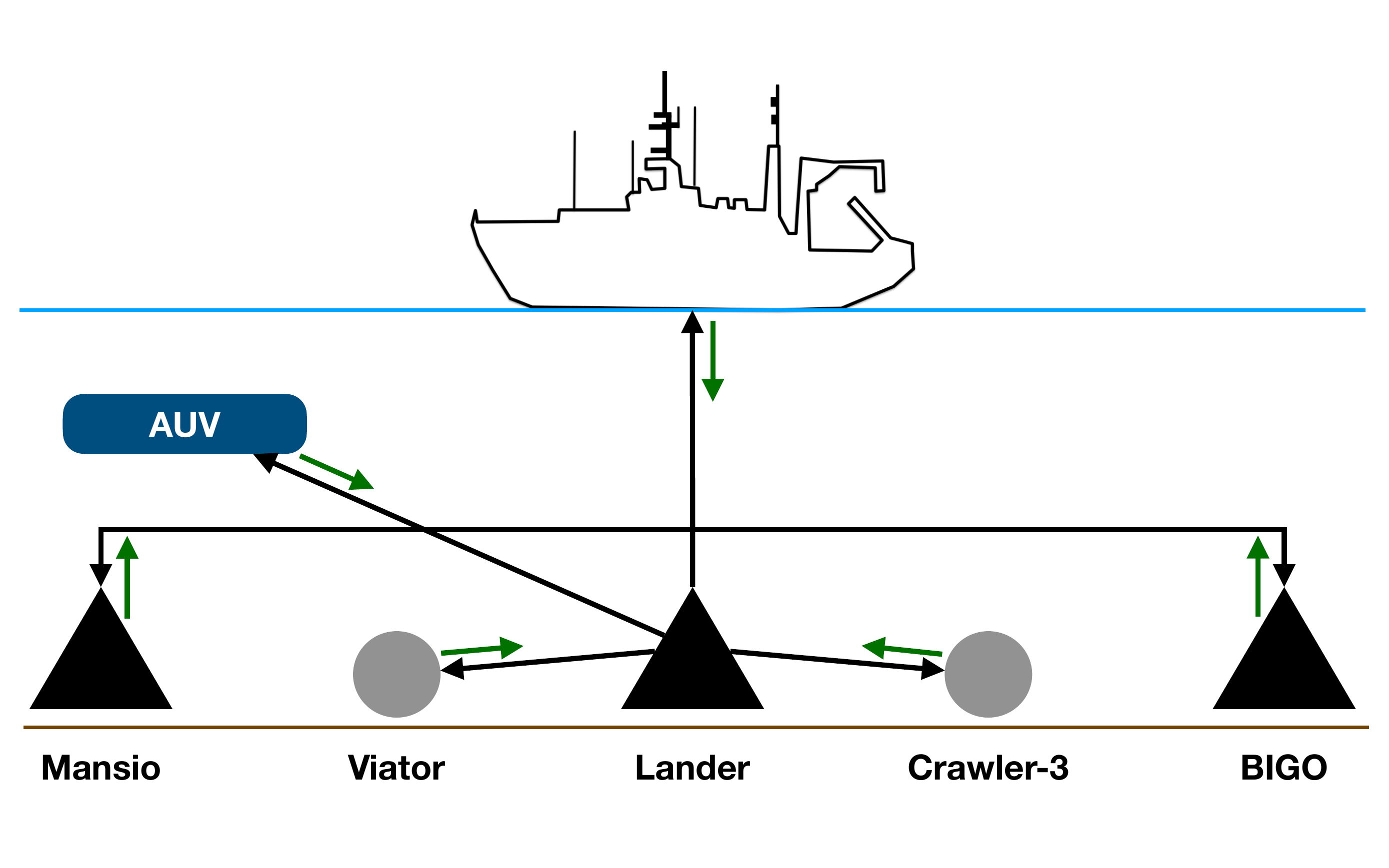}
  \caption{One platform detects an event and notifies the others}
  \label{fig:scenariod}
\end{subfigure}
\caption{Scenarios to develop the digital thread}
\label{fig:scenarios}
\end{figure*}

As mentioned in \autoref{sec:relatedwork} the goal of the project ARCHES is the implementation of an underwater network of collaborating OOSs. The prototyping method described in \autoref{subsec:dtp} was used to develop DTPs of all ocean observation systems that are part of the underwater acoustic network. Employing DTPs enables us to easily develop and test the scenarios presented in \autoref{fig:scenarios}. \autoref{fig:scenarioa} illustrates the basic scenario. To simplify it, each physical twin uses one of its sensors to measure oxygen and sends the results to the research vessel, where the digital twin receives the data.
Although the OOSs are planed to collaborate in the future, we still want to be able to operate them from a research vessel. Hence, in a second scenario (see \autoref{fig:scenariob}) commands are send to each OOS, which then respond with a confirmation including the current status of the OOS.
Since computational power underwater is limited by the energy supply of the OOSs, gathered data will be processed on the research vessel. On the research vessel, gathered data can be combined with external data such as weather data to predict upcoming events, e.g., storms. If such an event is predicted, commands are send to all participating OOSs, which adapt their measurement strategy according to a predefined protocol. Reconfiguration of the network includes changes in the sampling rate of various sensors, could trigger flux measurements by the lander, or movements of the mobile platforms to different positions. This scenario is presented in \autoref{fig:scenarioc}. \autoref{fig:scenariod} illustrates a collaboration scenario, where one OOS predicts an event and sends this to all the other OOSs. All OOSs then adapt their behavior to best possibly sample the expected event.

\subsection{Along the Life Cycle of RAMI 4.0}\label{subsec:dtptest}
 This approach advocates to follow the paradigms of \emph{Continuous Delivery} \emph{(CD)} \cite{humble2010continuous} to support a continuous software engineering process. This way, software engineering and embedded systems techniques are combined to ensure a high standard in software quality.

\subsubsection{Type (Development, Maintenance)}
\begin{figure}[tb]
  \centering
  \includegraphics[width=0.3\textwidth]{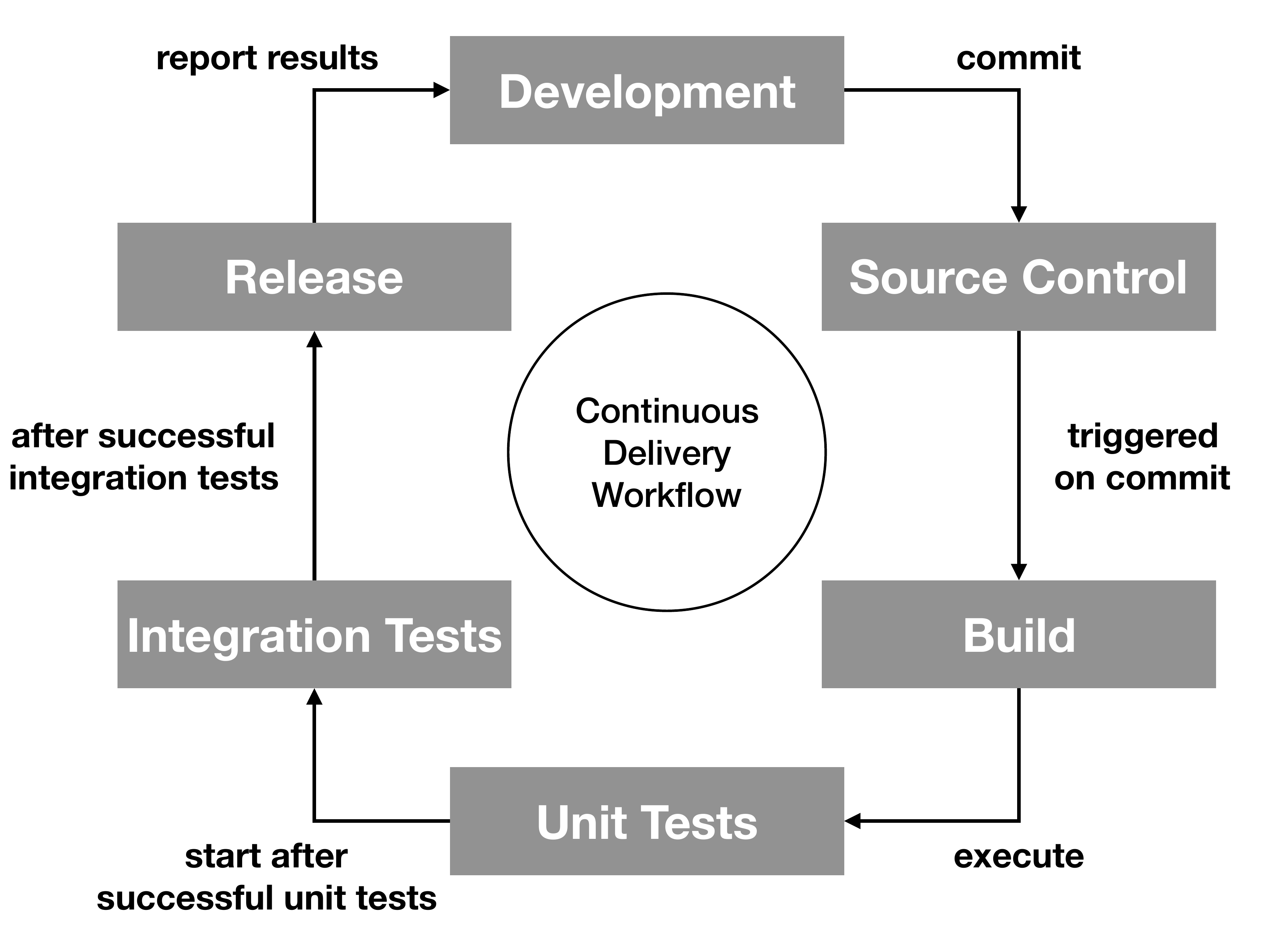}
  \caption{Continuous Delivery for Digital Twins}
  \label{fig:cd}
\end{figure}

The entire code base is managed in GitLab with a connected Docker runner on a virtual machine and a private Docker registry. Additionally, we connected GitLab runners installed on a RaspberryPI 3B+ and a RaspberryPI 4B (4GB RAM) to build/release ARM images and test the software on the hardware used in production. Each microservice is placed in a separate project. As shown in \autoref{fig:cd}, commits automatically trigger the build of the microservice inside a Docker container. Afterwards, the functional unit tests and integration tests are executed in this container. For the integration tests the test methods provided by ROS are used. These integration tests start the ROS nodes and execute tests on the running systems. At this point only parts of the DTPs with the emulated sensors and actuators are used, i.e., automated Software-in-the-Loop (SiL) tests are performed. To increase the ground truth of the tests, existing data from past missions are used. The advantage of this approach is that developers can test, evaluate, and even simulate extreme conditions in their software before the corresponding physical objects are deployed, e.g., underwater. This saves time and costs in developing and testing OOS and reduces risks. After all tests were successfully executed a Docker container is released. This workflow is executed on each branch. Thus, containers are released for development branches as well as for the ``master'' branch.

\subsection{Along the Hierarchy Levels of RAMI 4.0}\label{subsec:hierarchy}
One of the key requirements in RAMI 4.0 is the Administration Shell (AS) for each physical asset. Several assets can form a thematic unit with a common AS \cite{ramizvei}.

\subsubsection{Control Device}\label{subsubsec:dtpcontrol}
All scenarios in \autoref{fig:scenarios} are planed around a research vessel to monitor and control the OOSs on the seafloor. On the research vessel each OOS has a corresponding DT running, although the states cannot be synchronized in real-time. Nevertheless, each DT provides its own Administration Shell to monitor and operate the sensors/actuators. This Administration Shell is a local web-service encapsulated in a Docker container and uses a bridge from ROS to the service. To develop this Administration Shell, the different DTPs with emulated sensors and existing data sets from previous missions are used. The received data is plotted in the AS using the tool OceanTea \cite{oceantea}. To operate all OOSs in one place, each DT registers its Administration Shell with a central Administration Shell. At this point, the different DTPs are, again, very useful to incrementally develop and test the different parts of the Administration Shells.

\section{THREATS TO VALIDITY}\label{sec:threats}
In the establishment and operation of an underwater acoustic network we face some crucial limitations such as limited energy and a small bandwidth in the acoustic communication. Thus, it is not possible to synchronize the physical twin with the digital twin in real-time. Furthermore, not all data recorded by each OOS can be transmitted.

Another obstacle is the emulation of complex actuators. For example, while turning light on and off is quite simple, actuators such as a milling drill are quite complex to emulate. The emulation of the in- and output is trivial, yet the behavior during drilling is not easy to emulate.

Due to the performed SiL tests in this approach, the correctness of the interaction with physical parts depends on the correctness of the implementations in the emulators. Since it is impossible to guarantee one-hundred-percent safety, the software still has to be manually tested on the final hardware before putting the OOS into the sea. With HiL tests, this process could be automated, too. However, OOSs are expensive and we cannot afford to park OOSs solely for automated HiL tests.

First results in the CI/CD pipeline showed that some ROS packages consume too much RAM while building on a RPI 3B+. As a consequence, each RPI 3B+ uses a swapfile of 1GB now. In the ARCHES demo mission a RaspberryPI 4B with $4\,$GB RAM will be mounted on another Lander system. The performance differences will be evaluated and presented, afterwards.

\section{CONCLUSION AND FUTURE WORK}\label{sec:conclusion}
The RAMI 4.0 covers many important aspects of smart systems and the advocated service-orientation architecture supports an agile development process. The DTP concept picks up at the idea of agile development and we sketched an incremental approach to cover different layers along the axis of the model. Except for the acoustic modem emulator by Evologics, only open-source tools were used in the presented approach. The continuous delivery workflow ensures that prototypes are properly tested and manages the different releases of software packages, automatically.

The next step is the connection of our software prototypes with the corresponding CAD models of the OOSs. Additionally, a simulation tool like Gazebo \cite{gazebohp} can be included in the automated test pipeline to test the positioning of moving OOSs.

The final demonstration of our framework for the project ARCHES will be conducted in Boknis Eck near the Eckernf\"orde Bight (Western Baltic Sea) in Germany. During this demo mission different OOSs are placed on the ground of the Baltic Sea for twelve days. The demo mission aims to prove our approach under real conditions and evaluates the scenarios presented in \autoref{fig:scenarios}. The software of all sensing platforms is based on tools described in this paper. The precise configuration of the underwater acoustic network will be published elsewhere.

\section*{Acknowledgment}
The project is supported through the HGF-Alliance ARCHES - Autonomous Robotic Networks to Help Modern Societies and the Helmholtz Association.

\IEEEtriggeratref{5}
\bibliographystyle{IEEEtran}
\bibliography{IEEEabrv, DigitalTwinPrototype}

\end{document}